\documentclass[a4paper,11pt]{article}
\usepackage{pos}

\title{The Short Baseline Neutrino Program at Fermilab}
\ShortTitle{The SBN Program at Fermilab}

\author*[a,1]{Maurizio Bonesini} 

\affiliation[a]{Sezione INFN Milano Bicocca, Dipartimento di Fisica G. Occhialini \\
  Universit\'a di Milano Bicocca, Piazza Scienza 3, Milano, Italy}

\note{on behalf of SBN (Icarus, SBND, MicroBooNe collaborations)}

\emailAdd{maurizio.bonesini AT mib.infn.it}

\abstract{
The current status of the Short Baseline Neutrino (SBN)
project at Fermilab is reviewed.
While the installation of SBND is still in progress, ICARUS has taken 
its first neutrino data on beam: using 
both the Booster Neutrino Beam (BNB) and the Neutrino at the Main Injector 
(NuMI) beam. 

MicroBooNE has presently completed its data taking and is producing the
world's first high statistics results on $\nu$-Ar interactions, in both 
inclusive and exclusive channels. In parallel, the unexpected MiniBooNE 
``low energy excess'' is under investigation, to search for sterile
neutrinos. 

The physics potential  for 
sterile neutrino searches at SBN  will be outlined, with emphasis on the 
Neutrino-4 experiment and the possible ICARUS verification of this claim.}

\FullConference{%
  *** The 22nd International Workshop on Neutrinos from Accelerators (NuFact2021) ***\\
  *** 6–11 Sep 2021 ***\\
  *** Cagliari, Italy ***}

\begin{document}
\maketitle

\section{Introduction}
Anomalies observed in short baseline (SBL) oscillation experiments in the
last 20 years (see table \ref{tab-exp}), not fitting inside the 
standard model of 3-flavour neutrino mixing, may be explained by a 3+1 
oscillation model with a sterile neutrino, where $\Delta m^{2} \sim 
O(1 eV^{2})$ and sin$^{2} (2 \theta)$ is relatively small. Unluckily no model
has so far been successful in fitting all experimental results at once.
\begin{table}[htbp]
\smallskip
\centering
\begin{tabular}{|l|c|c|c|c|c|}
\hline
	Experiment   & type            & baseline, E$_{\nu}$ &   mode   &  channel & CL  \\
\hline
	LSND \cite{lsnd}        & DAR accelerator & $\sim 30$ m, $\sim 30$  MeV        &  appearance & $ \overline{\nu}_{\mu} \mapsto \overline{\nu}_{e}$  &  3.8 $\sigma$ \\
	MiniBooNE  \cite{miniboone}   & SBL accelerator & $\sim 540$ m, $\geq 500$ MeV       & appearance  & $\nu_{\mu} \mapsto \nu_{e} $                         & 4.5 $\sigma$ \\
	     &                 &                                   &             & $\overline{\nu}_{\mu} \mapsto \overline{\nu}_{e} $   & 2.8 $\sigma$ \\
GALLEX/SAGE \cite{gallex}  & source - e capture & 1.9 m, 0.6 m             &  disappearance & $\nu_{e} \mapsto \nu_{X} $ &  2.8 $\sigma $ \\	
	  &                 &  0.8 MeV      &    & &              \\
	Reactors  \cite{giunti}    & $\beta$ decay   &                           & disappearance & $\overline{\nu}_{e} \mapsto \overline{\nu}_{X}$  &  3.0 
	$\sigma $ \\ 
\hline
\end{tabular}
\caption{Main short-baseline experiments showing anomalies in the neutrino sector}
\label{tab-exp}
\end{table}

\begin{figure}[htbp]
\begin{center}
\includegraphics[width=\textwidth]{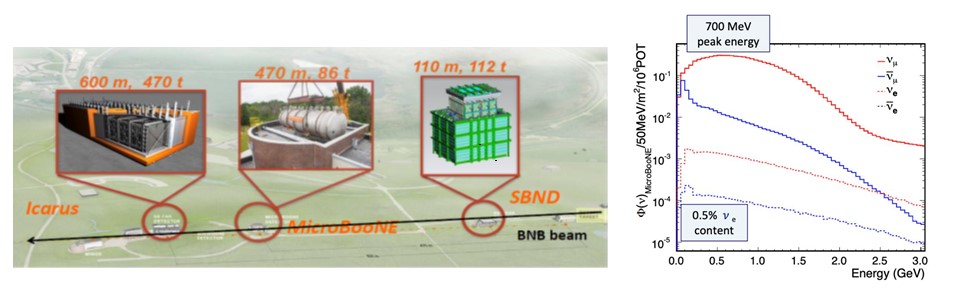}
\end{center}
\caption{Left panel: layout of the BNB beamline at Fermilab. For the SBN 
program SBND will operate as near detector and Icarus as Far detector. Right
panel: components of the BNB beam at Fermilab.}
\label{fig-beamline}
\end{figure}

The Fermilab Short-Baseline Neutrino program (SBN) \cite{sbn} has been
proposed in 2015 to give a definitive answer to the problem.
The SBN program will have the unique possibility to exploit both $\nu_e$ 
appearance and $\nu_{\mu}$ disappearance using the well known Booster Neutrino
Beam (BNB) 
\footnote{ fluxes are well understood thanks to a detailed simulation \cite{bnb}
and the availability of 8.9 GeV/c p+Be data from the HARP experiment at
CERN PS \cite{harp} }
and employing three detectors based on the common technology
of Liquid Argon (LAr) TPC.
As shown in figure \ref{fig-beamline}, SBND is located at 100 m from the BNB 
target, followed by MicroBooNE at 470 m and Icarus at 600 m. 
BNB is a conventional horn-focussed $\nu_{\mu}$ beam with an energy spectrum
peaking at $\sim 700$ MeV and with a well-known $\nu_{e}$ contamination
($\sim 0.5 \%$), see the right panel of figure \ref{fig-beamline}. 
While SBND will collect events from BNB at a rate of 0.25 Hz, Icarus will have
a 0.03 Hz rate. Being at ground level, with a limited overburden, both detectors
will have a sizeable rate of cosmics: around 0.03 Hz at SBND and 0.14 Hz 
at Icarus. The mitigation of this issue is one fundamental aspect of data taking
for both experiments. 
\section{The SBN Physics Program}
\begin{figure} [htbp]
\begin{center}
\includegraphics[width=\textwidth]{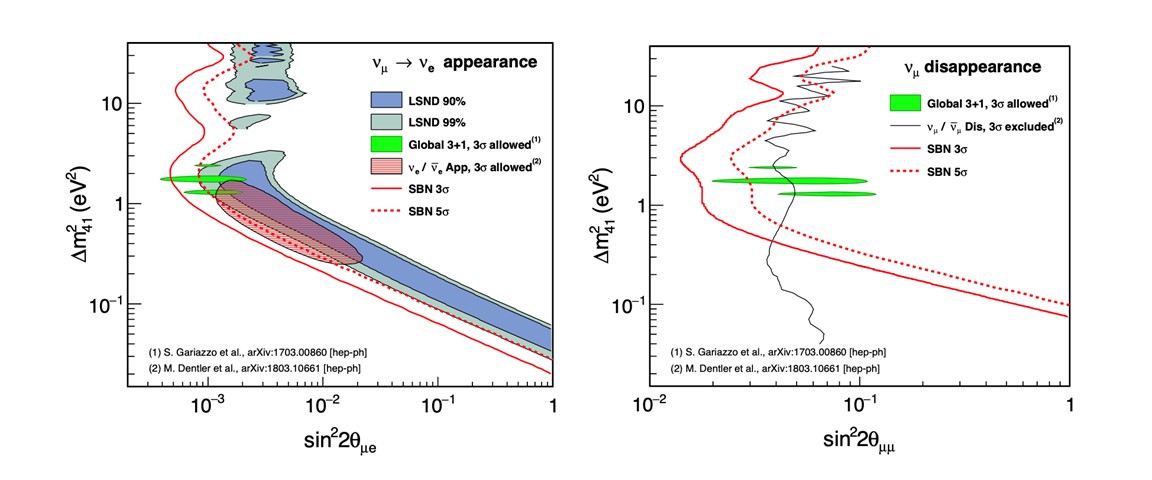}
\end{center}
\vskip -1 cm
\caption{Expected sensitivities in the SBN program, in $\nu_e$ 
appearance (left) and $\nu_{\mu}$ disappearance (right) modes, 
from reference \cite{sbn}}
\label{fig-sens}
\end{figure}

The SBN Physics Program was set up:
\begin{itemize}
\vskip -1.4 cm
\item{} to understand the nature of  MiniBooNE ``low energy excess'', 
with MicroBooNE (Phase I);
\vskip -1.4 cm
\item{} to search search for sterile neutrinos both in appearance and 
disappearance
channels, using SBND as near and ICARUS as far detector (Phase II).
The use of the same detector technology will greatly reduce the systematics  
errors, while the good $\nu_{e}$ identification capability of a 
LAr TPC will help to reduce backgrounds.
\vskip -1.4 cm
\item{} pave the ground for future long-baseline experiments, as DUNE 
\cite{dune}, by a further development of the LAr TPC technology and by a
high-statistics measure of $\nu -$Ar cross sections in the few GeV region.
\end{itemize}
SBND will accumulate $\sim 7 \times 10^{6}$ $\nu_{\mu}$ and $5 \times 10^{4}$ 
$\nu_{e}$ 
events in a 3-year run, to measure $\nu-$Ar cross sections in the DUNE 
energy range. ICARUS will collect $\sim 10^{5}$ events/year from the NuMI 
beam (off-axis) producing high statistics electron neutrino cross sections. 
Five sigmas sensitivities will be reached in the SBN program with 3-year data 
taking ($6.6 \times 10^{20}$ POT), to search for
sterile neutrinos in appearance or disappearance channels, in the
currently allowed parameter region. The expected sensitivities
are shown in figure \ref{fig-sens}.
In addition, a rich program of BSM searches will  be performed, 
looking  for  neutrino
tridents, dark matter, heavy leptons, Lorentz and CPT violations ...

\begin{figure} [htbp]
\begin{center}
\includegraphics[width=\textwidth]{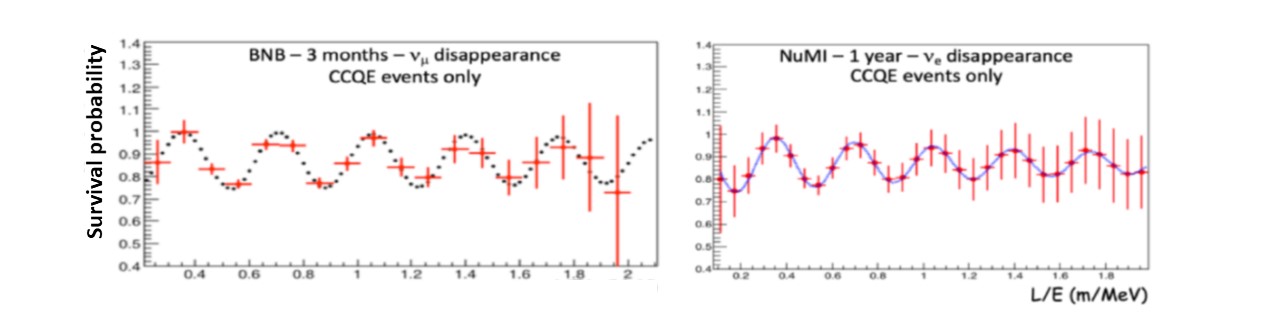}
\end{center}
\vskip -1 cm
\caption{
Left(right) panel: survival $\nu_{\mu}$ ($\nu_e$) probability for the 
Neutrino-4 anomaly and expected ICARUS measurement for 3 months (1 year)
of BNB (NuMI) data taking.}
\label{fig-neu4}
\end{figure}

The reactor Neutrino-4 experiment has recently shown evidence at 2.7 $\sigma$
for an oscillatory pattern with best fit parameters $\Delta m^{2}_{N4}=
7.30$ eV$^2$, sin$^2 \theta_{N4}=0.36$.
\cite{nu4}. Having a similar L/E ratio, ICARUS alone may provide a complete 
verification of this claim 
in less than one year, by using both the BNB and the NuMI beam \cite{rubbia1}.
About 11500 $\nu_{\mu}$ CC events will be collected in 3 months of
data taking at BNB and about 5200 $\nu_{e}$ interactions will be accumulated
in 1 year of data taking at NuMI. Survival $\nu_{\mu} (\nu_{e}$) probability
are shown in figure \ref{fig-neu4}, for the oscillation case.
Neutrino-4 expectations are shown as black dotted line or
blue continous line in the figure. 	

\section{Status of the SBN Program}
The SBN detectors are based on the LAr TPC technology introduced
by C. Rubbia in 1977 \cite{rubbia} and developed by the ICARUS Collaboration
in the following years  \cite{icarus}. 

\begin{figure} [htbp]
\begin{center}
\includegraphics[width=\textwidth]{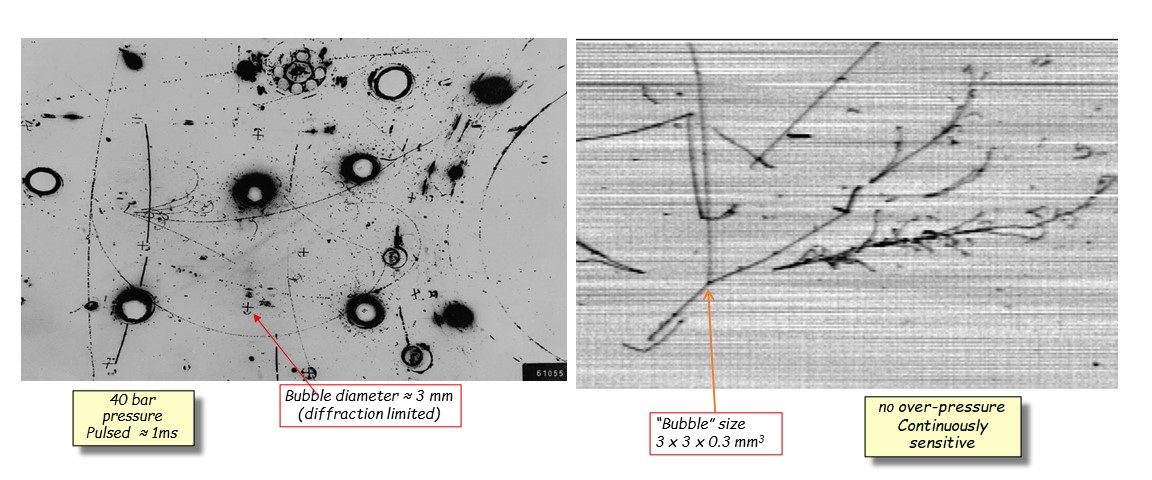}
\end{center}
\vskip -0.7 cm
\caption{Images of two neutrino events as seen by the giant heavy freon bubble
chamber ``Gargamelle'' at CERN (left side) and by ICARUS T600 (right side). 
``Bubble'' sizes are similar.}
\label{fig-lar}
\end{figure}

A LAr TPC is a kind of ``electronic bubble chamber'' that gives detailed 
images as shown in figure \ref{fig-lar}.
As a figure of merit in the heavy freon bubble chamber 
``Gargamelle'' the sensitive mass was 3 tons, compared
to the 600 tons of ICARUS LAr TPC. 
\subsection{Phase I: MicroBooNe}
MicroBooNE is currently the world's longest running LAr  TPC: from
2015 on. With a mass of 170 tons of liquid Argon (87 tons active volume), it 
consists of a TPC with 2.5 m drift,
a system of 32 8'' PMTs coated with TPB and a top/side cosmic ray tagger (CRT)
to reject cosmics. 
It has presently completed its physics run, with more than 33 published papers.
It has produced the world's first high statistics precision studies of $\nu$-Ar 
interactions, in both inclusive and exclusive final states \cite{mistry}. 
In addition, it is investigating the unexpected MiniBooNE excess 
($ > 3 \sigma$ statistical and systematics combined) at lower energies 
of neutrino interactions producing final
states electrons or photons.
The two possibilities may not be discriminated by a Cherenkov detector as
MiniBooNE, in contrast with MicroBooNE that uses a different
technology and thus has the possibility to assess if these events are due to
$\nu_{e}$ charged current interactions, see reference \cite{sutton} for 
more details. 
Unfortunately, due to its surface location, MicroBooNE has very limited
capabilities in non-accelerator $\nu$ physics.
\subsection{Phase II: SBND+Icarus}
\begin{figure} [htbp]
\begin{center}
\includegraphics[width=\textwidth]{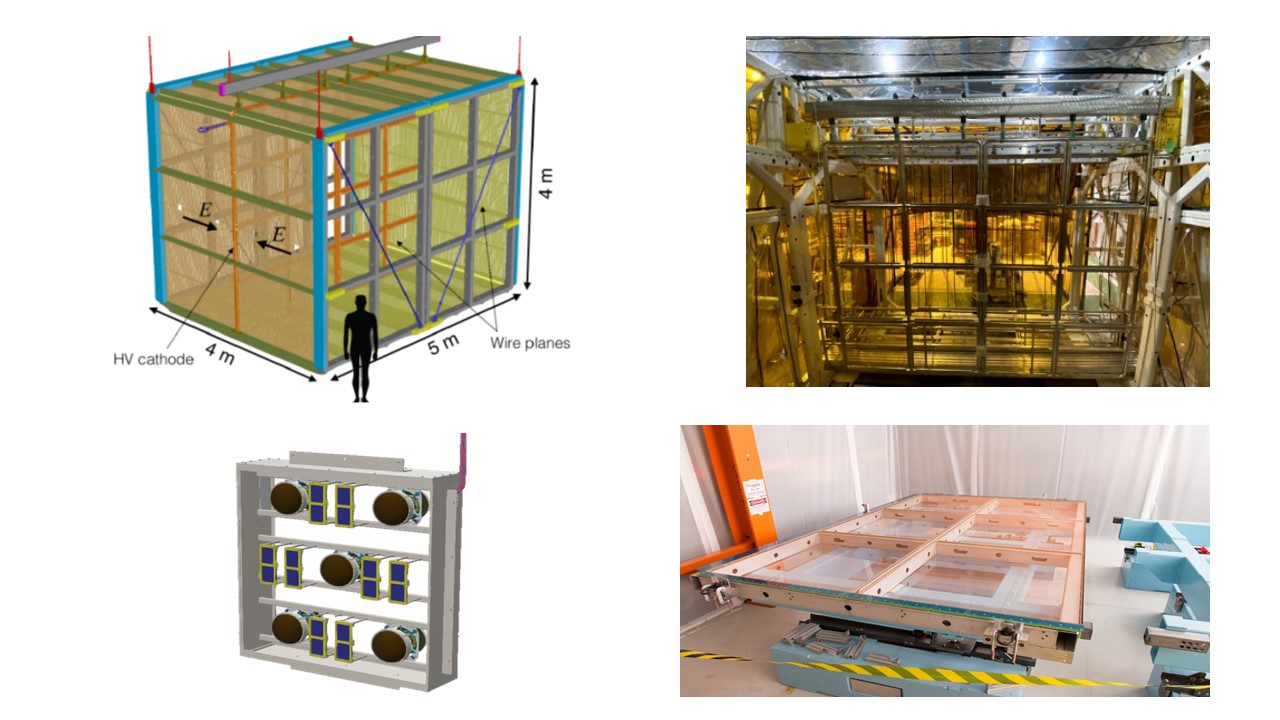}
\end{center}
\vskip -0.7 cm
	\caption{Left panel: layout of the SBND TPCs (top), layout of one
PDS module (bottom). Rght panel: cathode plane
assembly (CPA) recently installed (top); first APA plane on-site
(bottom)}
\label{fig-sbnd}
\end{figure}

Phase II of the SBN program will use both SBND as near detector and ICARUS 
as far 
detector to study SBL neutrino oscillations.
With a mass of 260 tons of LAr (112 tons active mass) SBND 
is made of 2 TPCs with 2m drift, a 
light detection system based on 120 8" PMTs (96 coated with TPB) and 192 
X-ARAPUCA photon traps instrumented with SiPMs and a $4 \pi$ CRT. 
One central cathode plane assembly (CPA) divides the TPC active volume (
$5 \times 5 \times 4 m^3$) in two drift volumes. The readout is based on two
anode planes assemblies (APA), made of 3 wire planes each (vertical, 
$\pm 60^{\circ}$) with a wire pitch $\sim 3$ mm, see figure \ref{fig-sbnd}
for details. 
Behind APAs, 24 photon detection modules (PDS) are placed. Each is made of 
5 PMTs and 8 X-ARAPUCA modules.
Every side of the detector will be covered by planes of extruded scintillator
strips, making the CRT. 
In regards to the installation at Fermilab, PMTs and X-ARAPUCAs have been tested
and delivered, TPC installation is under way as well as 
CRT and cryostat/cryogenics and
will be finished by 2022.

\begin{figure}
\begin{center}
\vskip -0.8 cm
\includegraphics[width=.65\textwidth]{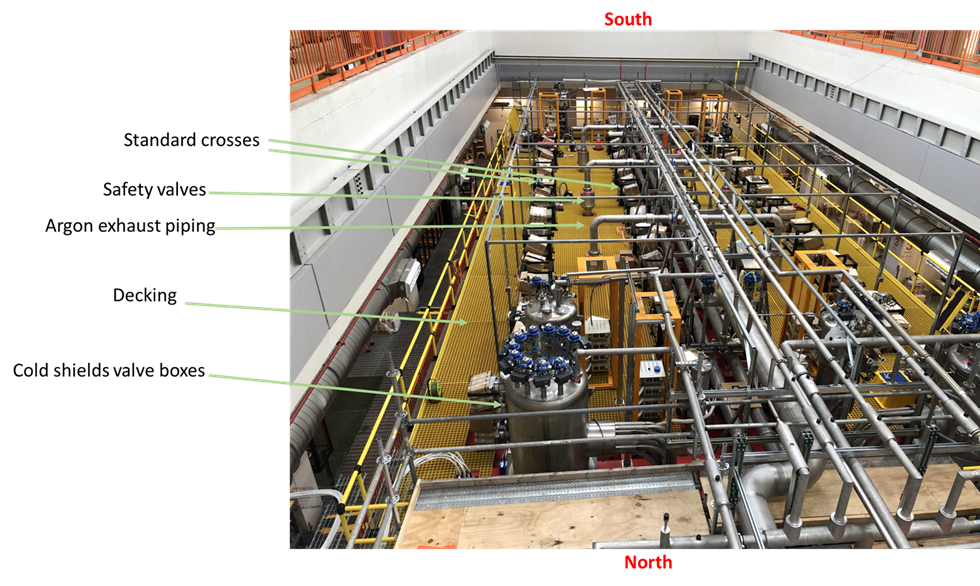}
\includegraphics[width=.34\textwidth]{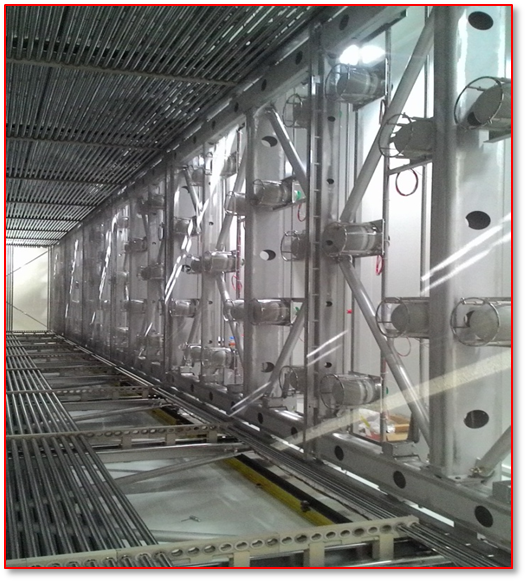}
\end{center}
\vskip -0.7 cm
\caption{Left: layout of the cryogenics for ICARUS at Fermilab. 
Right: PMTs after the planes of readout wires.
For more details see \cite{torti}.}
\label{fig-inst}
\end{figure}

ICARUS has an active mass of 
476 tons of liquid Argon. It was refurbished at CERN in the framework of
the WA104 program \cite{wa104}.
Cold vessels and a purely passive insulation were installed; a new scintillation
light detection system (based on 360 PMTs coated with TPB) was provided; a new
TPC readout electronics was implemented and cryogenics and LAr purification
systems were refurbished.
The present installation status of the ICARUS cryogenic system at Fermilab,
on the top of the detector, is shown in
figure \ref{fig-inst}.
Aside the top CRT, installation activities in ICARUS are completed, with latest
adjustments during COVID-19 restricted operations. Since March 17th 2020
the detector is operated with 24/7 remote shifts and a minimal presence in
situ. 
One of the main problems in ICARUS data-taking is its installation at shallow depth,
with a limited overburden. ICARUS will thus be exposed to a continuos flux of
cosmic rays, containing high-energy muons that may be misidentified as part of
a neutrino interaction. To mitigate this problem, two handles are available:
a $4 \pi$  CRT surrounding ICARUS TPCs and the timing in coincidence with the beam spill,
from the PMTs system. 
The 3m overburden, on the top of ICARUS, will absorb more than $99 \%$ of the 
incoming photons and hadrons. The CRT system is 
composed of a bottom, a side and a top CRT. 
Each is made of two layers of plastic scintillators, partly recuperated
from Double CHOOZ and MINOS. The CRT will provide informations on incoming 
muons by their detection and measurement of crossing time and coordinates. 
Preliminary results are shown in reference \cite{bishu}.

The new light detection system \cite{pmt} is based on 360 8" R5912-MOD photomultipliers,
installed behind the TPCs wires, as shown in figure \ref{fig-inst}. The PMTs' 
coating with TPB allows the detection of the prompt scintillation light from 
LAr at 128 nm.
The system will allow ICARUS to identify the time of occurrence ($t_0$) of any 
ionizing event in the TPC with $\sim 1$ ns timing resolution, localize events 
with $\leq 50$ cm spatial resolution and determine their rough topology and
generate a trigger signal for readout.
Together with the beam bunched structure, this will allow an efficient
rejection of the cosmics background. PMTs'  equalization of gain and timing 
is performed with a laser system, flashed on each PMT by a dedicated fiber
system \cite{laser}.

\begin{figure} [htbp]
\vskip -0.8 cm
\begin{center}
\includegraphics[width=0.9\textwidth]{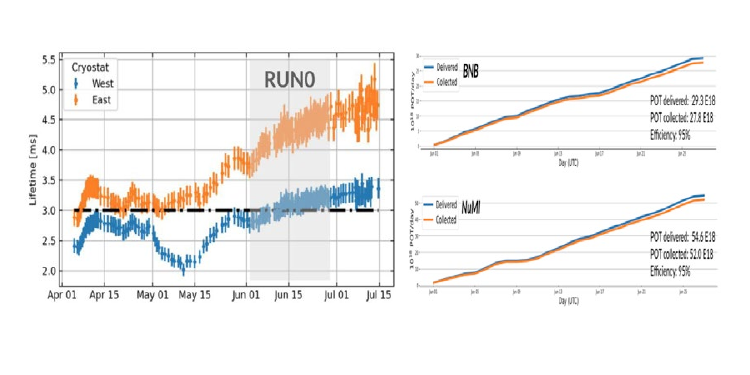}
\end{center}
\vskip -1.5 cm
\caption{Left panel: measured electron drift lifetime during ICARUS RUN0.
Right panel: delivered and collected data during RUN0 from the ICARUS detector.}
\label{fig-run}
\end{figure}
The other major issue in ICARUS data taking is the Argon purity level, that affects
the electron drift lefeime $\tau_{D}$. As shown in figure \ref{fig-run} in RUN0
$\tau_{D}$ reached up to $\sim 4.5 (3)$ ns in the EAST (WEST) cryostat, allowing efficient 
signal detection over the full LAr volume.  This was due to improvements on 
poor performing GAr recirculation units and periodic venting (3 times/day). 
Further improvements are expected from the installation of new higher capacity
GAr filters.
\section{ICARUS RUN0 with BNB and NuMI beamlines}
\begin{figure} [htbp]
\begin{center}
\includegraphics[width=\textwidth]{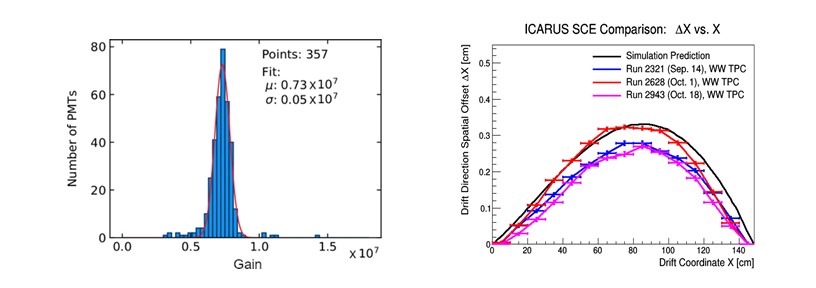}
\end{center}
\vskip -1.1 cm
\caption{Left panel: Distribution of PMTs' gains after the equalization 
	procedure. Right  panel: measurement of SCE for the ICARUS TPC. 
	}
\label{fig-res}
\end{figure}

During RUN0, up to end of July 2021, ICARUS has taken data with the
following goals:
\begin{itemize}
\item{} certify the readiness of the detector for physics quality data, 
operating as primary BNB user;
\item{} verify the possibility to run the detector in remote mode 24/7,
	with a limited on-site presence;
\item{} test DAQ for different triggers for both BNB and NuMI beam;
\item{} accumulate data samples of good quality to tune neutrino and cosmics
event reconstruction.
\end{itemize}
Two main triggers we used to collect BNB and NuMI events:
a Minimum Bias Trigger, where data were recorded for every beam spill and
a ``Majority trigger'' using pairs of discriminated PMT signals in 
coincidence with the beam spill gate.
As shown in figure \ref{fig-run} about $27.8 \times 10^{18} (52.0 \times 
10^{18}$ POT were collected from BNB (NuMI) beams with an efficiency 
$\sim 95 \%$. 
As examples of the commissioning of ICARUS in RUN0, figure \ref{fig-res} 
shows the results of the PMTS' gain equalization using the laser system 
and single photoelectrons from the background~\footnote{ after equalization 
the spread of PMTs' gains is reduced to $\sim 7 \%$ from an initial value
$\sim 17 \%$, obtained using preliminary calibrations at room  temperature} 
and
of the measurement of space charge effects (SCE) using anode-cathode-crossing 
cosmic muons, looking at distortions in the drift direction. 

\begin{figure} [htbp]
\begin{center} 
\includegraphics[width=0.8\textwidth]{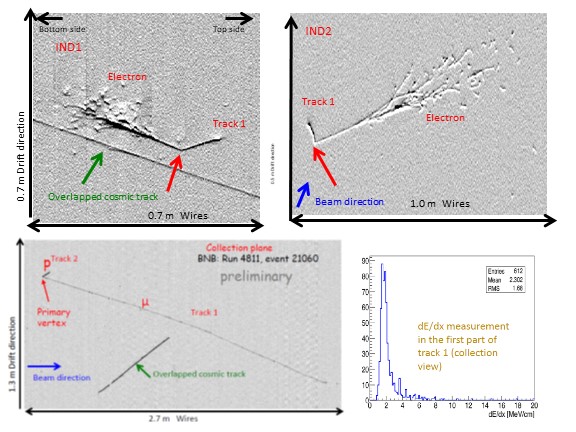}
\end{center}
\vskip -0.7 cm
\caption{Examples of CC QE neutrino events, collected during RUN0 at Fermilab.
	Top: NuMI $\nu_{e}$ CC candidate; bottom: BNB contained $\nu_{\mu}$ 
	candidate. Red arrows point to the primary vertex.}
\label{fig-scan}
\end{figure}

The ICARUS  detector recorded its first muon and electron neutrinos, thus
demonstrating its excellent detection capabilities.
About 254 $\nu_{\mu}$ CC and 15 $\nu_{e}$ CC candidates were selected and 
visually scanned during this run. Examples of charge current (CC)
quasi-elastic (QE) candidates 
from BNB and NuMI beams are shown in figure
\ref{fig-scan}.
The top panels of figure \ref{fig-scan} show a picture of a NuMI QE CC
$\nu_{e} p \mapsto p e $ event. Track 1 is the forward going proton candidate
stopping inside 13 cm. The e.m. shower from the electron candidate ($E_{dep} 
\sim 650$ MeV) is going downward. 
The left bottom panel of figure \ref{fig-scan} shows instead the image of
a BNB CC QE $\nu_{\mu} n \mapsto p \mu$ event. Track 1 is the muon candidate 
stopping after 2.8 m, with a deposited energy $E_{dep} \sim 650$ MeV, while
track 2 is the proton candidate stopping after 10.9 cm, with $E_{dep} \sim
100$ MeV. The shower beginning is clearly visible.
\section{Conclusions}
The SBN program at Fermilab is progressing well, with MicroBOONE now 
producing high statistics measurements of $\nu-$Ar cross sections and
getting closer to release the first results on the low energy excess of
MiniBooNE, with $6 \times 10^{20}$ POT.

ICARUS was activated in August 2020 and is currently taking data.

Assembly and installation of SBND detector are progressing and will finish
by 2022.


\begin{thebibliography}{99}
\bibitem{lsnd} 
A.Aguilar-Arevelo {\it et al.} [LSND Coll.], `` Evidence for neutrino
oscillations from the observation of $\overline{\nu}_{e} $
appearance in a $\overline{\nu}_{\mu}$ beam'', Phys. ReV. D64 (2001) 112007.
\bibitem{miniboone} A. Aguilar-Arevelo {\it et al.} [MiniBooNE Coll.], 
	`` Improved Search for 
$\overline{\nu}_{\mu} \mapsto \overline{\nu}_{e}$ Oscillations in the MiniBooNe
		experiment'', Phys. ReV. Lett. 110 (2013) 161801.
\bibitem{gallex}
C. Giunti, M. Laveder, ``Statistical Significance of the Gallium Anomaly'',
		Phys. ReV. C83(2011) 065504.
\bibitem{giunti} C. Giunti {\it et al.}, ``Update of short-baseline electron
	neutrino and antineutrino disappearance'', Phys. ReV. D86 (2012) 113014

\bibitem{sbn}
R. Acciari et al., ``A Proposal for a Three Detector Short-Baseline Neutrino Oscillation Program in the Fermilab Booster Neutrino Beam'', 
arXiv:hep.ins-det/1503.01520.

\bibitem{bnb} 
A. Aguilar-Arevado {\it et al.} [MiniBooNE Coll.], ``Neutrino
flux prediction at MiniBooNE'', Phys. ReV. D79 (2009) 072002.

\bibitem{harp} M.G. Catanesi {\it et al.} [HARP Coll] 
`` Measurement of the production cross-section of positive pions in the 
collision of 8.9 GeV/c protons on beryllium '', Eur. Phys. J. C52 (2007) 20
\bibitem{dune} 
B. Abi {\it et al.} [DUNE Coll.] ``Long-baseline neutrino oscillation physics 
potential of the DUNE experiment'', Eur. Phys. J. C80 (2020) 10978.

\bibitem{nu4} 
A.P. Serebrov,R.M. Samoilov, 
``Analysis of the results of the Neutrino-4 experiment 
on the search for sterile neutrino and comparison with results of other
		experiments'', JETP Lett. 112 (2020) 4; \\
A.P. Serebrov {\it et al.} [Neutrino-4 Coll], ``First Observation of the 
Oscillation Effect in the Neutrino-4 Experiment on the Search for Sterile
		Neutrno'', JETP Lett. 109 (2019) 213.
\bibitem{rubbia1} C. Rubbia ``Experimental searches of neutrino anomalies'', 
talk at NEUTEL 2021.
\bibitem{rubbia} C. Rubbia {\it et al.}, ``The Liquid Argon Time Projection
Chamber: A New Concept In Neutrino Detector'', CERN-EP/77-08 1977.

\bibitem{icarus}
M. Antonello {\it et al.} [ICARUS Coll.], ``Design, construction and tests
of the Icarus T600 detector'', Nucl. Instr. Meth. {\bf A527} (2004) 329.

\bibitem{mistry} K. Mistry, ``Recent Neutrino Cross Section Results from 
MicroBooNE'', talk at Neutel 2021
\bibitem{sutton} K. Sutton, ``MicroBooNE's Search for a Photon-Like Low 
Energy Excess'', these proceedings

\bibitem{wa104} M. Bonesini [ICARUS Coll.] ``THe WA104 experiment at CERN'', 
J. Phys. Conf. Ser 650 (2015) 1, 012015.

\bibitem{torti} M. Torti, ``Short-baseline neutrino oscillation searches 
with the ICARUS detector'', these proceedings 

\bibitem{bishu} B. Behera, ``First data from the commissioned ICARUS side 
cosmic ray tagger'', these proceedings.

\bibitem{pmt}
B. Ali-Mohammadzadet {\it et al.}, ``Design and implementation of the new scintillator 
light detection system of ICARUS T600'', JINST 15 (2020) T10007.
\bibitem{laser}
M. Bonesini {\it et al.}, `` The development of the ICARUS T600 laser diode
calibration system'', Nucl. Instr. Meth. A936 (2019) 261. 

\end{thebibliography}
\end{document}